\documentclass[aps,prl,twocolumn]{revtex4}

\usepackage{graphicx,latexsym,color}

\def\be{\begin{equation}}

\def\ee{\end{equation}}

\def\bea{\begin{eqnarray}}

\def\eea{\end{eqnarray}}

\def\bi{\begin{itemize}}

\def\ei{\end{itemize}}

\newcommand{\comment}[1]{}

\begin{document}

\title{Analysis of Localization Phenomena in Weakly Interacting Disordered Lattice Gases}

\author{T. Schulte$^1$, S. Drenkelforth$^1$, J. Kruse$^1$, R. Tiemeyer$^1$, K. Sacha$^{2,3}$,
J. Zakrzewski$^{2,3}$, M. Lewenstein$^{4,5,*}$, W. Ertmer$^1$, and J.J. Arlt$^1$}

\affiliation{$^1$ Institut f\"ur Quantenoptik, Leibniz Universit\"at
Hannover, Welfengarten 1, D-30167 Hannover, Germany}

\affiliation{$^2$Instytut Fizyki imienia Mariana Smoluchowskiego,
Uniwersytet Jagiello\'nski, PL-30-059 Krakow, Poland}

\affiliation{$^3$ Marc Kac Complex Systems Research Center,
Jagiellonian University, PL-30-059 Krakow, Poland}

\affiliation{$^4$ Institut f\"ur Theoretische Physik, Leibniz
Universit\"at Hannover, D-30167 Hannover, Germany}

\affiliation{$^5$ ICFO - Institut de Ci\'encies Fot\'oniques,
08034 Barcelona, Spain}

\begin{abstract}
Disorder plays a crucial role in many systems particularly in solid
state physics. However, the disorder in a particular system can
usually not be chosen or controlled. We show that the unique control
available for ultracold atomic gases may be used for the production
and observation of disordered quantum degenerate gases. A detailed
analysis of localization effects for two possible realizations of a
disordered potential is presented. In a theoretical analysis clear
localization effects are observed when a superlattice is used to
provide a quasiperiodic disorder. The effects of localization are
analyzed by investigating the superfluid fraction and the
localization length within the system. The theoretical analysis in
this paper paves a clear path for the future observation of
Anderson-like localization in disordered quantum gases.
\end{abstract}

\maketitle

In condensed matter or statistical physics disorder is typically
neither avoidable nor controllable. It is known to lead to
spectacular phenomena, such as Anderson localization
\cite{pwa,anderson}. In this case the interference of waves
scattered on random impurities or defects is responsible for
localization. In the so called weak localization regime coherent
backscattering may be regarded as a precursor for Anderson
localization \cite{Kramer}. For a recent treatment of light waves
scattered by atomic ensembles see e.g. \cite{miniat}.

Cold atomic gases in magnetic and optical potentials offer the
unique possibility to introduce well controlled disorder to the
system. For this purpose several methods have been proposed to
create disordered, or quasi-disordered potentials. These proposals
include the use of speckle radiation \cite{dainty, bodzio, Kuhn},
incommensurable optical lattices \cite{superlattices, Roth, Roth2,
bodzio, Davidson} and the interaction with impurity atoms \cite{Castin}. The
unique control provided by these methods should enable novel studies
of disorder induced effects, inaccessible to other systems. In
addition disorder appears also naturally close to the surface of
atom chips \cite{chiprough} and leads to a fragmentation of
Bose-Einstein condensates (BEC).

The first experimental attempts to systematically study the effects
of disorder on BEC used a speckle potential. The Florence group
\cite{inguscio} observed the apparent fragmentation of the
condensate into pieces in the presence of such a random potential,
resulting in characteristic stripes in the time-of-flight density
profiles. Another experiment, performed simultaneously in Florence
\cite{fort} and in Orsay \cite{aspect} showed that in the presence
of the random potential the expansion of the condensate is strongly
inhibited. A plausible scenario for this suppression of transport
given in \cite{aspect,aspect2,modugno} notes that the transport
stops when the BEC encounters a random potential modulation of
sufficient height. This explanation is purely classical but it is
supported by simulations of the Gross-Pitaevskii equation (GPE).

Another disorder configuration was realized in our group by
combining a one dimensional optical lattice and a random potential
\cite{wir}. This potential configuration was designed to mimic the
situation first envisaged by Anderson \cite{anderson}. However, it
was shown that, contrary to first expectations, one can not observe
Anderson localization in this experiment due to interaction effects
and the length scale of the employed disorder potential. We discuss
this situation in more detail in the following sections.

The onset of the Bose-glass phase has recently been observed
\cite{fallani06} for ultracold atoms in the strongly interacting
regime as suggested by \cite{bodzio}. For more details on strongly
interacting bosonic systems we refer the reader to the seminal
discussion in \cite{fisher}. We also note that mixtures of fermions
and bosons in random optical potentials are of considerable interest
\cite{anna}. One component in such a mixture may in fact serve as
the disorder in the system \cite{Sengstock}. Here we concentrate
solely on the weakly interacting case for bosons and investigate the
circumstances for the appearance of Anderson-like localization.

The paper is organized as follows. In the first part we describe the
experimental realization of a weakly interacting lattice gas with a
superimposed {\it slowly} varying disorder potential and discuss the
ground state properties of such a system. We observe a classical
fragmentation of the condensate which is confirmed by a theoretical
analysis based on the GPE but no signature of Anderson localization
is found.

Therefore we theoretically investigate another type of disorder
which has a shorter correlation length in the second part. Following
earlier suggestions \cite{Roth,Roth2,bodzio} we consider the ground
state of the condensate in the presence of a pseudorandom potential
introduced by additional optical lattices of different wavelengths.
Such a potential is called a superlattice \cite{bodzio, Roth,
Roth2}. We discuss the influence of nonlinear interactions on the
occurrence of localization phenomena in this system.

\section{Disordered lattice potential}

\subsection{Experimental results}

Despite a variety of possible realizations of disorder, one method
is particularly straightforward to implement in current experimental
setups. It consists of projecting a disordered optical dipole
potential onto the atomic sample. Following the theoretical
suggestion \cite{bodzio} such realizations have recently been used
to investigate the effects of disorder on the ground state
\cite{wir,inguscio} and on the dynamics of weakly interacting BECs
\cite{inguscio, fort, aspect}.

We perform our experiments with $^{87}$Rb Bose-Einstein condensates
in an elongated magnetic trap. After laser cooling and trapping cold
atoms are loaded into a cloverleaf magnetic trap with axial and
radial frequencies of $\omega_x=2\pi\times 14$~Hz and
$\omega_\perp=2\pi\times 200$~Hz, respectively. These atoms are
evaporatively cooled to quantum degeneracy resulting in a final
number of condensed atoms between $1.5\cdot10^4$ and $8\cdot10^4$.

To carry out experiments in a disordered lattice configuration we
use two optical dipole potentials, a 1D lattice and the disordered
potential. Both are derived from a Ti:Sa laser operating at a
wavelength of $\lambda=825$~nm. Acousto-optic modulators are used to
control the intensity of each dipole potential and optical fibres
are employed to deliver the light beams to the experiment.

The optical lattice is created by retro-reflection of a laser beam
along the axial direction of the magnetic trap. The depth of the
optical lattice is typically set to 6.5~$E_r$. The recoil energy
$E_r$ is given by $E_r=\hbar^2 k^2/2m$, where $m$ denotes the atomic
mass and $k$ corresponds to the wave number of the optical lattice.
The detection system is used to image the axial position of the
atomic cloud as well as the the beam waist. This allows for precise
positioning of the beam waist with respect to the atomic sample.

\begin{figure}[h]
\centering
\includegraphics*[width=8.6cm]{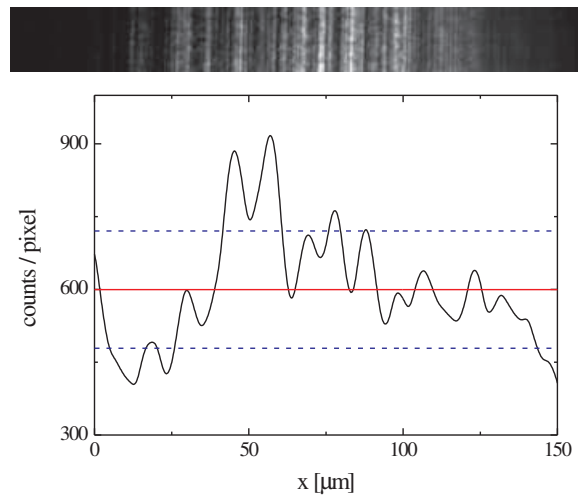}
\caption{Typical intensity distribution of the disordered optical
dipole potential. An image of the intensity distribution is shown at
the top. The intensity variation in a small region corresponding to
the size of the sample is shown below.} \label{speckle}
\end{figure}

\begin{figure}[h]
\centering
\includegraphics*[width=8.6cm]{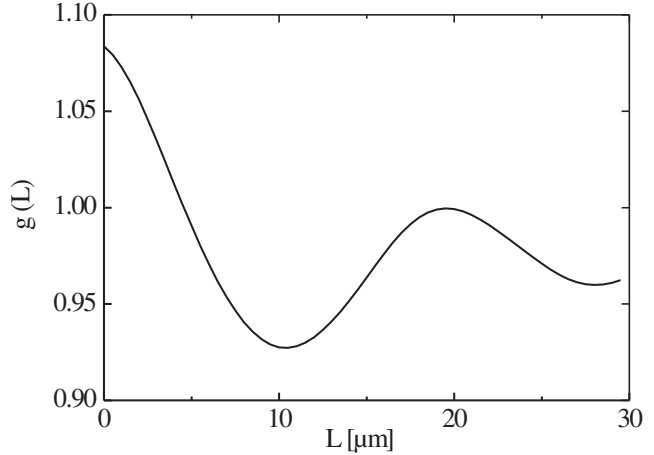}
\caption{Auto correlation function for a typical realization of the
disordered potential.} \label{correlation}
\end{figure}

The disorder potential is created by illuminating a randomly
structured chrome substrate with a laser beam. The partially
transmitting substrate is imaged onto the atomic sample from a
direction orthogonal to the optical lattice. Due to the resolution
of the imaging system, the minimal structure size of the dipole
potential is limited to a few microns. Figure~\ref{speckle} shows an
image of the disorder potential at the position of the atomic cloud.
We define the depth of the disorder potential $V_{\Delta}$ as twice
the standard deviation of the dipole potential, analogously to
\cite{inguscio}. Figure \ref{correlation} shows the calculated auto
correlation function

\begin{equation}
g(L)=\frac{\left\langle
V_{dis}(x)V_{dis}(x+L)\right\rangle}{\left\langle
V_{dis}(x)\right\rangle \left\langle V_{dis}(x+L)\right\rangle}
\label{autocorrelation}
\end{equation}

for a typical realization of the disordered potential. In
(\ref{autocorrelation}) the brackets denote the average over
position and $V_{dis}(x)$ represents the disordered potential.
Figure \ref{correlation} shows that the correlation length of the
potential decays on a typical length scale of $10\; \mu m$.

After the production of the BEC our experiments are performed as
follows. The optical lattice is ramped to its final depth within
60~ms. Subsequently the disorder potential is increased to its final
depth within another 60~ms followed by a hold time of 20~ms. Finally
all potentials are switched off and the atomic density distribution
is measured after 20~ms of free expansion using absorption imaging.

\begin{figure}
\centering
\includegraphics*[width=8.6cm]{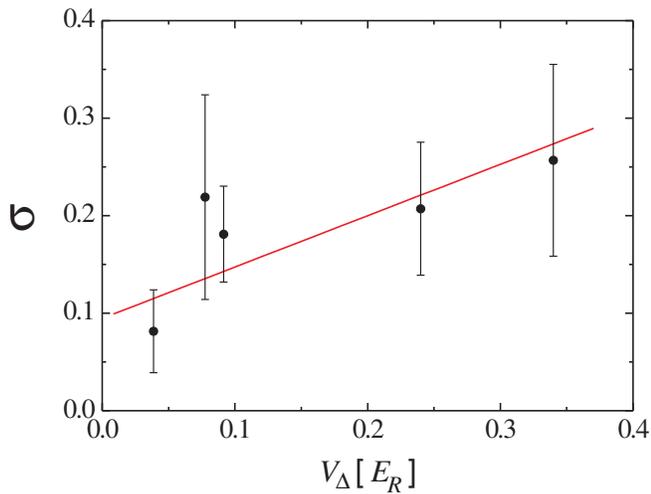}
\caption{Standard deviation $\sigma$ of the observed axial density
distribution from an inverted parabola after 20~ms of free
expansion, depending on the disorder strength. The red line is a
linear fit to the data which serves as a guide to the eye.}
\label{Res_vs_Dis}
\end{figure}

In the expanded atomic density distributions we observe significant
irregular modulations which depend on the strength of the disordered
potential. To quantify these modulations we fit the central peak of
the expanded lattice gas with an inverted parabolic distribution and
calculate the standard deviation $\sigma$ of the measured density
profile from this fit. Figure \ref{Res_vs_Dis} shows the dependence
of the standard deviation on the depth of the disordered potential.
Clearly the deviation from the unperturbed parabolic distribution
grows with increased disorder strength.

\begin{figure}
\centering
\includegraphics*[width=8.6cm]{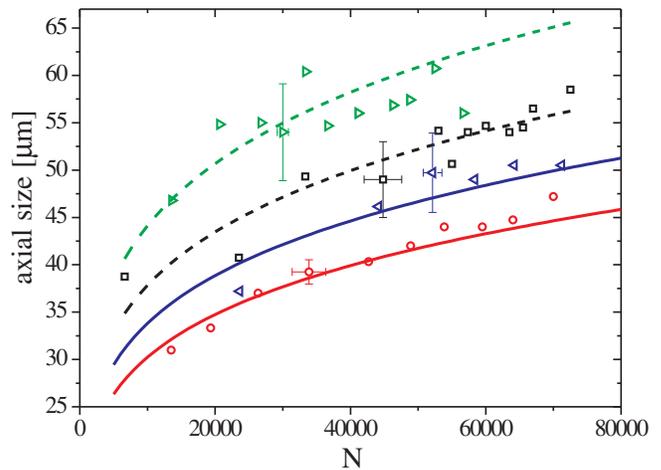}
\caption{Size of the central peak after 20~ms of free expansion
versus the number of atoms. The clouds are released from the
following potentials: magnetic trap (red \textcolor{red}{$\circ$}),
magnetic trap and disorder (black $\Box$), magnetic trap and optical
lattice (blue \textcolor{blue}{$\lhd$}), magnetic trap and disorder
and lattice (green \textcolor{green}{$\rhd$}). The lines correspond
to a theoretical prediction (see text). The lattice depth is
6.5~$E_r$ and the disorder has a depth of 0.1~$E_r$.} \label{expres}
\end{figure}

To further investigate the expansion of the disordered lattice gas
we extract the axial size of the central momentum peak from the fits
of the density with an inverted parabolic distribution.
Figure~\ref{expres} shows the resulting sizes for four different
configurations of the combined potential as a function of the atom
number.

The red curve shows the theoretical prediction based on the
Thomas-Fermi approximation and subsequent self-similar expansion
\cite{CastinDum} for a confinement in the magnetic trap only. The
blue curve shows an estimate for the expansion of a lattice gas in
the absence of disorder. It is obtained by using the increased axial
size of the cloud due to the presence of the optical lattice
\cite{Pedri} as a starting point for the self similar expansion.

The addition of the disorder potential yields a more surprising
result. Despite its very small potential depth, the additional axial
confinement due to the disorder potential leads to a significant
increase of the axial size after expansion. The theoretical analysis
below shows that the disordered potential induces strong deviations
of the density profile from the parabolic envelope. Therefore
significant deviations from the self similar expansion are expected
for disordered gases. According to figure \ref{Res_vs_Dis} these
deviations produce irregular density modulations and can lead to
pronounced changes in the widths of the expanded clouds. The
explicit form of the expanded density distribution however depends
strongly on the exact realization of the disordered potential. We
have used a 3D numerical simulation to check the expansion of the
disordered gas in the absence of the optical lattice. Depending on
the disorder potential used, the simulation confirms the observed
behavior qualitatively.

\subsection{Theoretical Analysis}

In the system considered here, the disorder potential is imposed
along one axis of the cloud. Hence a 1D Gross-Pitaevskii equation is
used to describe the basic properties of the system and to analyze
parameter regimes where localization phenomena can be observed. The
GPE provides an appropriate description of the system as long as
depletion effects are small. We have estimated the number of atoms
depleted from the condensate wave-function within the Bogoliubov
theory for the optical lattice potential used in the experiment. For
a total particle number of $10^5$ atoms the fraction of depleted
atoms is less than 1\%.

The 1D GPE used in our simulations is given by

\be
i\partial_t\phi=\left[-\frac{\partial^2_x}{2}+\frac{x^2}{2}+V_0\cos^2(kx)+V_{\rm
dis}(x)+g|\phi|^2\right]\phi, \label{gpe} \ee

where we have adopted harmonic oscillator units, i.e.
$\hbar\omega_x$, $\sqrt{\hbar/m\omega_x}$ and $1/\omega_x$ as
energy, length and time units. The depth of the optical lattice is
given by $V_0$ and $V_{\rm dis}(x)$ denotes the disordered
potential. In the 3D case the coupling constant is given by $g_{\rm
3D}=4\pi \hbar^2 a/m$, where $a$ denotes the $s$-wave scattering
length. For the 1D simulations the coupling constant $g$ is chosen
such that the Thomas-Fermi radius of the 1D gas equals the axial
Thomas-Fermi radius in the 3D trap.

To simulate the experimental situation the wavelength of the lattice
is set to $\lambda=825\ nm$. Since the harmonic trap and the
disorder potential change on a length scale much greater than the
lattice spacing and the condensate healing length, $l=1/\sqrt{8\pi n
a}$ (where $n$ is the condensate density) we eliminate the lattice
potential in the GPE by applying the so called effective mass
analysis \cite{effmass}.

\begin{figure}
\centering
\includegraphics*[width=8.6cm]{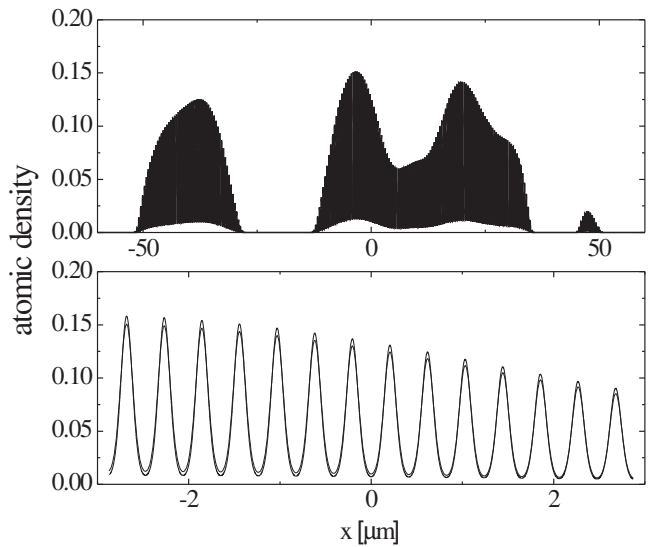}
\caption{Top panel: Ground state solution of the GPE for a
condensate in the combined potential of the magnetic trap
($\omega_x= 2\pi \times 14$~Hz), optical lattice and disorder
potential. Bottom panel: Central part of the ground state solution
of the GPE (solid line) and the corresponding solution within the
effective mass approach (dashed line). The depth of the optical
lattice is 6.5$\;E_r$ while the depth of the disorder potential
$0.7\;E_r$.} \label{effmass}
\end{figure}

Within this analysis the GPE~(\ref{gpe}) is replaced by an equation
where the optical lattice potential is absent but the mass of a
particle and the coupling constant are corrected. We assume that the
ground state solution of the GPE has the form

\be \phi_0(x)=\sqrt{\cal N}f(x)u_0(x). \label{ans} \ee

Here $u_0(x)$ denotes the Bloch function corresponding to the ground
state of the Schr\"odinger equation with the optical lattice
potential, $f(x)$ is an envelope function and $\cal N$ is a
normalization factor. Substituting the ansatz (\ref{ans}) into
Eq.~(\ref{gpe}) leads to the equation

\be \mu^* f(x)=\left[-\frac{\partial^2_x}{2m^*}+\frac{x^2}{2}+V_{\rm
dis}(x)+g^*|f(x)|^2\right]f(x), \label{effe} \ee

where $m^*$ and $g^*$ are the effective mass and the effective
coupling constant, respectively. For an optical lattice depth of
6.5~$E_r$ the effective parameters are

\bea m^*=2.56\;m, \cr g^*=1.66\;g. \eea

For the total number of $N=10^5$ atoms (that implies $g=1800$) the
values of the effective parameters suggest that one can employ the
Thomas-Fermi approximation and neglect the kinetic energy term. Then
the solution of Eq.~(\ref{effe}) is

\be |f(x)|^2=\frac{\mu^*-x^2/2-V_{\rm dis}(x)}{g^*}, \ee

where $\mu^*$ is determined from the normalization condition \be
\int |f(x)|^2{\rm d}x=1. \ee In Fig.~\ref{effmass} we show a
comparison of the ground state solution of the full GPE (\ref{gpe})
and the solution of the effective mass approach obtained within the
Thomas-Fermi approximation. The squared overlap of these solutions
is greater than 0.99 and hence they are practically identical.

This analysis shows that the effect of a slowly varying disorder
potential applied to Bose-Einstein condensates in the lattice
potential can be described within the Thomas-Fermi approximation.
The condensate density is modulated by the slowly varying disorder
and consequently no Anderson localization is present.

These results thus bear similarity to the experiments performed in
the absence of an optical lattice \cite{inguscio} where a
fragmentation of the BEC is induced by the disorder potential.

\section{Superlattice potential}

The above analysis suggests that it is necessary to introduce a
disorder that changes on a length scale smaller than the healing
length to enter a regime where localization effects can be observed.
To overcome the experimental difficulties of imposing such a truly
random potential the use of pseudorandom potentials has been
suggested \cite{Roth,bodzio}.

These pseudorandom potentials can be formed by one or more
additional optical lattices creating a so called superlattice. Since
cold atomic gases present a finite sized system, a suitably chosen
pseudorandom potential can provide the desired disorder
\cite{Sokoloff}.

\begin{figure}
\centering
\includegraphics*[width=8.6cm]{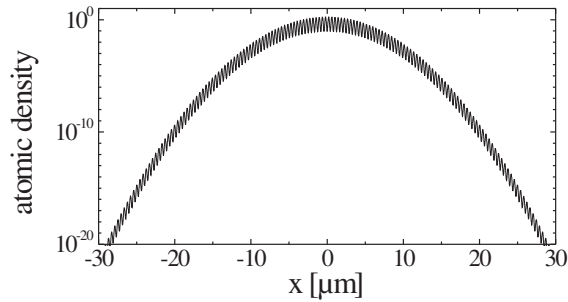}
\caption{Ground state density of a non-interacting BEC in the
combined potential of a magnetic trap ($\omega_x=2\pi \times 4\ Hz$)
and an optical lattice with a wavelength of $\lambda=825\ nm$ and a
depth of $6.5\ E_r$. Note the logarithmic vertical scale.}
\label{nloc}
\end{figure}

\begin{figure}
\centering
\includegraphics*[width=8.6cm]{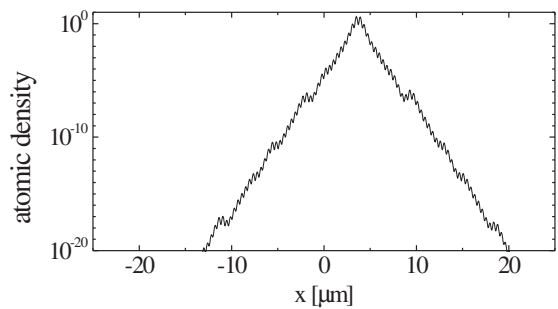}
\caption{Ground state of a non-interacting BEC in the potential of
Fig.\ref{nloc} with two additional weak optical lattices at
$\lambda=960\ nm$ and $\lambda=1060\ nm$ with depths of $0.2E_r$.}
\label{loc}
\end{figure}

Consider first a situation without interactions as discussed
previously  in \cite{bodzio}. Figure~\ref{nloc} shows the ground
state density for a condensate in the combined potential of the
magnetic trap and the optical lattice. As expected the modulation of
the density due to the lattice is visible. The addition of two very
weak lattices at wavelengths of $\lambda=960 nm$ and $\lambda=1060
 nm$ changes this situation drastically as shown in Fig.~\ref{loc}.
The exponential tails of the ensemble density are a clear
manifestation of Anderson-like localization.

However, it is well known that localization phenomena are strongly
influenced by the presence of interactions \cite{scalettar}.
Therefore a simulation including these interactions is necessary to
predict parameter regimes for the observation of localization
phenomena.

\subsection{Screening due to interaction}

\begin{figure}
\centering
\includegraphics*[width=8.6cm]{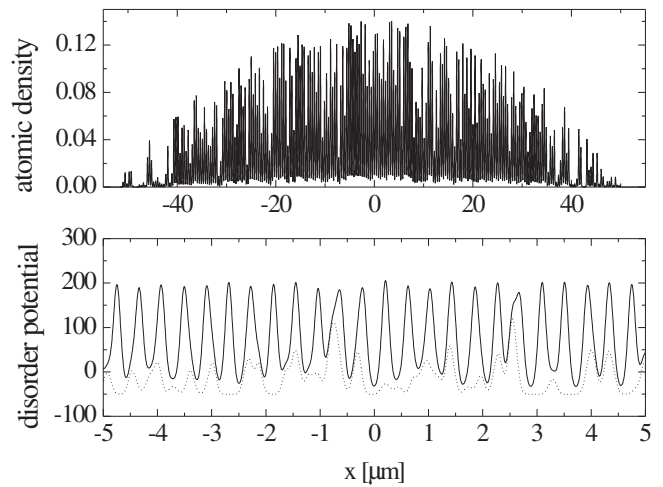}
\caption{Top panel: Ground state solution of the GPE for a
condensate in the combined potential of the magnetic trap, optical
lattice and disorder potential. Bottom panel: Disorder potential
$V_{\rm dis}(x)$ (dashed line) and disorder potential plus the
$g|\phi_0(x)|^2$ term (solid line). The depth of the optical lattice
is 6.5$\;E_r$ while the depth of the disorder potential is
$0.4\;E_r$. The coupling constant is $g=1800$.} \label{screen}
\end{figure}

Figure~\ref{screen} shows the ground state solution of the GPE for a
disordered potential that changes on the length scale of the optical
lattice. While the effective mass approach can no longer be applied
in this case, the localization signatures still do not emerge. This
is due to the fact that the interaction between atoms delocalize the
condensate.

The effect can be visualized by considering an effective potential.
That is, once the solution $\phi_0$ of the stationary GPE is known,
the equation can be considered as a Schr\"odinger equation

\be -\frac12\frac{\partial^2\phi_0(x)}{\partial x^2}+V_{\rm
eff}(x)\phi_0(x)=\mu\phi_0(x), \ee

with the effective potential

\be V_{\rm eff}(x)=\frac{x^2}{2}+V_0\cos^2(kx)+V_{\rm
dis}(x)+g|\phi_0(x)|^2.
\ee

As shown in Fig. \ref{loc} the situation without the nonlinear term
$g|\phi_0(x)|^2$ leads to the observation of Anderson-like
localization. However, when this term is present the effective
potential loses its disordered nature due to a screening effect.
Figure~\ref{screen} shows the potentials $V_{\rm dis}(x)$ and
$V_{\rm dis}(x)+g|\phi_0(x)|^2$. It is apparent that in the
effective potential the disorder is smoothed and screened. For
typical experimental parameters the term $g|\phi_0(x)|^2$ dominates
over $V_{\rm dis}(x)$ and consequently the randomness necessary for
localization is lost.

Hence a fine scale disorder alone is not sufficient to induce a
non-trivial localization in the system. Additional control of the
interaction is necessary. This effect is due to the accumulation of
atoms in the wells of the random potential. In these regions of high
density the nonlinear term in the GPE effectively smoothes the
potential modulations \cite{becdis}. A detailed discussion of this
screening effect was recently given in \cite{Sanchez}.

\subsection{Analysis of Localization}

We have investigated the transition from a fully delocalized to a
localized ground state of a bosonic lattice gas in two ways. First a
range of interaction strengths $g|\phi_0(x)|^2$ were investigated
numerically for a fixed depth of the additional lattices. In current
experiments a variation of the interaction strength can be realized
by varying the density $|\phi_0(x)|^2$ of the sample or by using a
Feshbach resonance to change the coupling constant $g$. Our main aim
was to investigate if suitable parameters for the experimental
observation of localization can be obtained. Secondly we have
investigated the onset of localization as the disorder strength is
increased in a weakly interacting BEC.

Our calculations were performed in an experimentally accessible
regime with a trap frequency of $2\pi\times 4$~Hz and a pseudorandom
potential equivalent to the one used for Fig.~\ref{loc}. For $g=0$
one obtains Anderson-like localization of the ground state
wavefunction which is characterized by an exponential localization
as shown in Fig~\ref{loc}.

\begin{figure}
\centering
\includegraphics*[width=8.6cm]{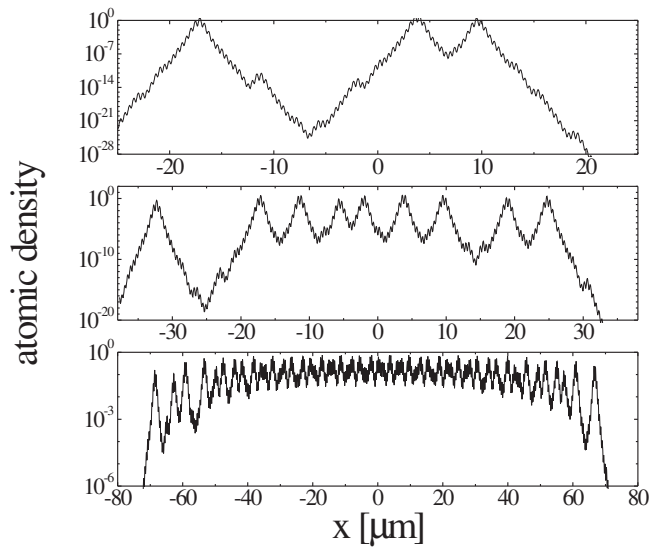}
\caption{Ground states of the GPE  for a condensate in the combined
potential of the magnetic trap, optical lattice and pseudorandom
potential. The depth of the optical lattice is 6.5$\;E_r$ while the
depths of the additional lattices forming the pseudorandom potential
are $0.2\;E_r$. The coupling constants $g$ for the panels are: 0.5
(top), 8 (middle), 256 (bottom). Note the different scales in the
panels.} \label{B}
\end{figure}

Figure~\ref{B} shows ground states within this potential for three
values of the interaction parameter. As $g$ is increased the number
of localization centers grows and for large values of $g$ they
overlap considerably. This behavior suggests that the condensate
wavefunction becomes a combination of these localized states due to
nonlinear interactions.
When $g$ is of the order of 500 one can no
longer distinguish individual localized states and the clear
signature of non-trivial localization vanishes.

The case of $g=256$ shown in Fig.~\ref{B} is equivalent to a 3D
experimental realization with trap frequencies of
$\omega_x=2\pi\times 4$~Hz and $\omega_\perp=2\pi\times 40$~Hz and
$N=10^4$ atoms. The ground state simulation shows that
characteristic features of Anderson-like localization are present
while the experimental parameters are within reach.

\begin{figure}
\centering
\includegraphics*[width=8.6cm]{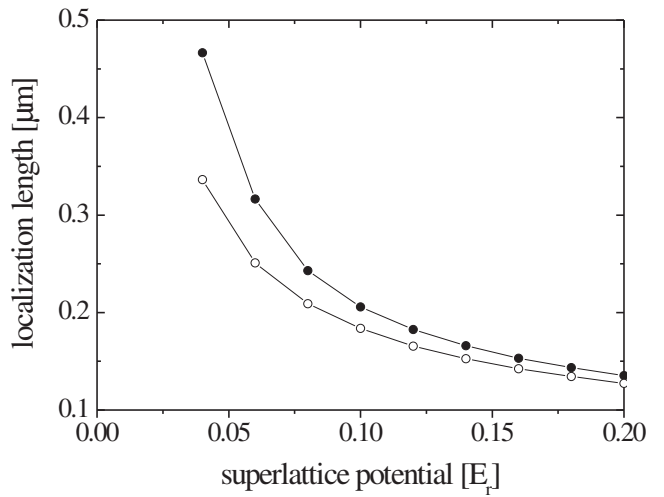}
\caption{Localization length as a function of the depth of the
additional optical lattice potentials for two interaction strengths
$g=0$ (circles) and $g=8$ (full dots). The depth of the main optical
lattice is 6.5$\;E_r$. } \label{loclength}
\end{figure}

In a second part of our analysis we have investigated the
localization length as a function of the disorder depth. The
localization length $l$ was obtained by fitting the individual
localization sites within the ground state density by
$|\phi_0(x)|^2\propto \exp(-|x-x_0|/l)$.

Figure~\ref{loclength} shows these localization lengths for depths
of the superlattice potential up to $0.2\ E_r$. The simulation was
performed for interaction strengths of $g=0$ and $g=8$ in a trap
with $\omega_x=2\pi\times4$~Hz trapping frequency. Each point in the
figure is the result of a fit to peaks in the wavefunctions depicted
in Fig.~\ref{B}. Within the scope of our simulation the amplitude of
the pseudorandom potential has to be sufficiently big to ensure that
the localization length is smaller than the size of the system.

This analysis clearly shows two distinct features of localization.
First, the localization length strongly depends on the depth of the
superlattice. Even a very small added pseudorandom potentials leads
to localization in small localization sites. This confirms the
expected non-perturbative character of localization. Secondly the
analysis shows that the interaction strength only has a small effect
on the localization length in the strongly localized regime as shown
in Fig.~\ref{loclength}. This can also be inferred from Fig.~\ref{B}
taking into account the different axis scales.

\subsection{Analysis of Superfluidity}

\begin{figure}
\centering
\includegraphics*[width=8.6cm]{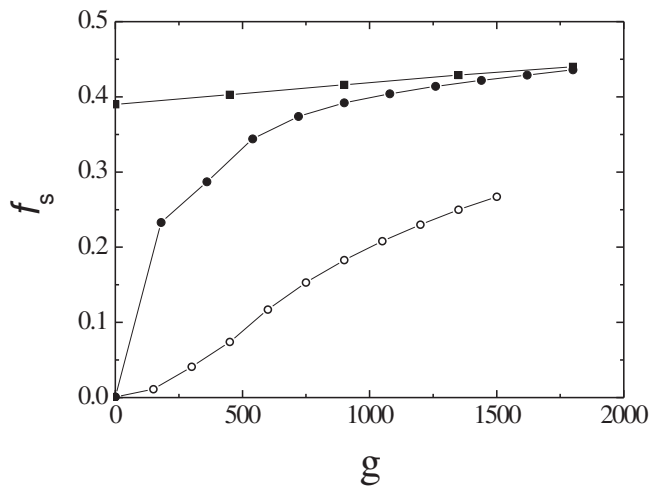}
\caption{Superfluid fraction as a function of the coupling constant
$g$ for a fixed superlattice depth of $0.2\ E_r$. Full dots are
obtained for a box potential corresponding to a trap frequency of
$\omega_x=2\pi\times14$~Hz and open circles to
$\omega_x=2\pi\times4$~Hz. Full squares show the case without a
superlattice potential for $\omega_x=2\pi\times14$~Hz.}
\label{theosf}
\end{figure}

To further investigate the effect of interactions we have analyzed
the superfluid fraction of the sample as a function of the coupling
constant $g$ in a potential box.

The superfluid fraction is obtained by calculating the response of
the condensed sample to twisted boundary conditions
\cite{superfluid}. Within this model the superfluid fraction
acquires additional kinetic energy due to a spatially varying phase.
A comparison of the energy with and without phase twist yields the
superfluid fraction which is defined as $f_s=2[E_0(v)-E_0(0)]/Nv^2$.
Here $E_0(v)$ is the ground state energy when a velocity field $v$
is imposed on the system (i.e. we compute the ground state solution
in the form $\phi_0(x)\exp(ivx)$ where $\phi_0(x)$ fulfills periodic
boundary conditions) \cite{superfluid}. In our calculations the size
of the potential box was chosen to match the size of the atomic
cloud in the harmonic potential. It is important to note that this
method only represents one possible definition of superfluidity
\cite{superfluid, twodefsuperfluid}. Within the GPE framework, it
does not include higher excitation modes or atoms depleted from the
condensate \cite{Roth2}.

Figure~\ref{theosf} shows the superfluid fraction as a function of
the coupling constant. In the presence of an optical lattice some
loss of the superfluid fraction is observed, but even for small
interaction strengths a considerable superfluid fraction remains.
This behavior drastically changes when the pseudorandom disorder
created by the two additional optical lattices at 960~nm and 1060~nm
is added. At low values of the coupling constant $g$ the superfluid
fraction strongly decreases, indicating the onset of localization.
However, our analysis shows that the superfluid fraction remains
large for the coupling constants $g$ in typical experimental
realizations, indicating the absence of Anderson-like localization.

\subsection{Detection of localization}

\begin{figure}
\centering
\includegraphics*[width=8.6cm]{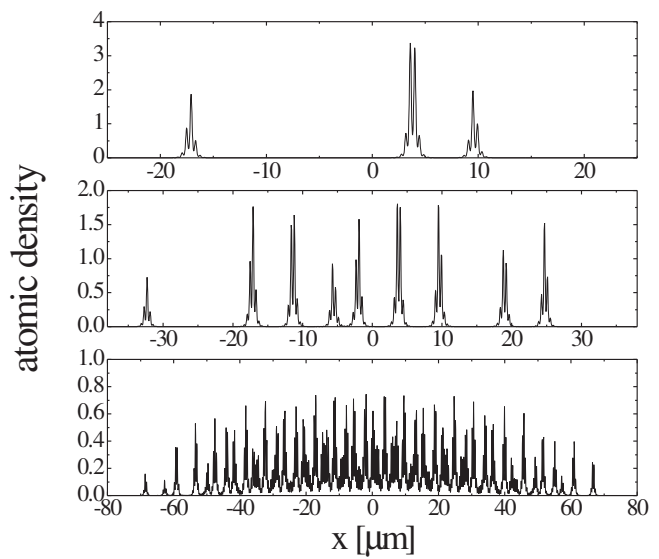}
\caption{Ground states of the GPE (previously shown in Fig.~\ref{B}
on a logarithmic scale) for a condensate in the combined potential
of the magnetic trap, optical lattice and pseudorandom potential.
The coupling constant $g$ for the panels are: 0.5 (top), 8 (middle),
256 (bottom).} \label{dens}
\end{figure}

The observation of features of Anderson-like localization may pose
considerable experimental difficulties. Figure~\ref{dens} shows the
density distributions on a linear scale. In an experimental
measurement the exponential nature of the density variation within
the localization sites will probably not be visible. Most likely,
limitations due to the imaging optics will inhibit the observation
of individual localization sites for experimentally accessible
densities.

However, the onset of localization leads to a considerable change in
the ground state density when a small pseudorandom potentials is
added. If the localization effect is indeed non-perturbative, it may
be possible to detect this change of the density even for small
added pseudorandom potentials.

\begin{figure}
\centering
\includegraphics*[width=8.6cm]{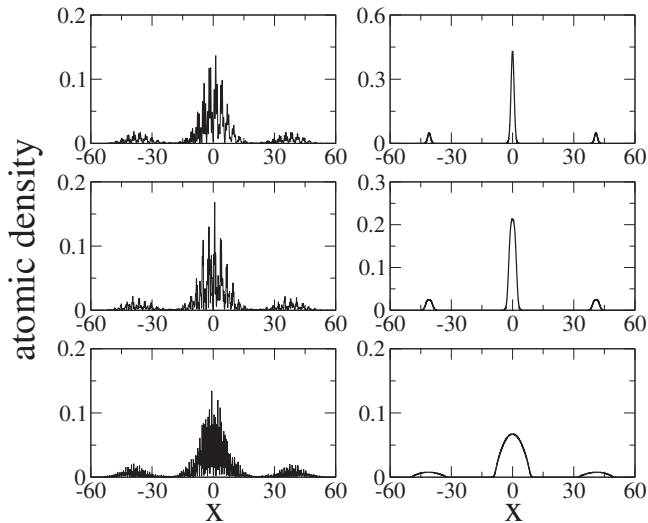}
\caption{Atomic density after 20~ms of free expansion for a
condensate prepared in the states shown in Fig.~\ref{dens} (left
column) and without disorder potential (right column). Oscillator
units corresponding to a trap frequency of $2\pi\times 4$~Hz are
used. } \label{C}
\end{figure}

A second avenue for the detection of localization is a
time-of-flight measurement of the velocity distribution. The
simulated density distribution after 20~ms of free expansion is
shown in Fig.~\ref{C}. Despite a clear difference in the ground
state wavefunction, the width of the envelope of the zero-momentum
peak is strikingly similar for all realizations. The results
represent a clear distinction from the classical case, where the
width of the expanded cloud depends strongly on the interaction
parameter $g$. We conclude that the width of the zero-momentum peak
mainly depends on the localization length $l$, which does not vary
significantly as a function of  $g$ in the simulations presented
here.

These two options show a path towards the observation of
Anderson-like localization in the regime discussed.

\section{Conclusion}

We have presented a detailed theoretical analysis of two
configurations for the production and observation of disordered
quantum degenerate gases. This analysis showed that localization
effects are not expected for the configurations involving a slowly
varying disordered dipole potential. However, clear localization
effects and a reduction of the superfluid fraction were observed
when a superlattice was used to provide the disorder. These effects
can be suppressed due to screening by nonlinear interactions within
the sample. The dependence of the localization features on these
interactions and on the depth of the superlattice potential were
analyzed in detail. It was shown that an analysis of the
time-of-flight signal will allow for a conclusive evaluation of
possible localization phenomena in the sample.

Within the experimental part a realization of a disordered lattice
gas is described in detail. An analysis of the time-of-flight signal
allows for a conclusive evaluation of possible localization
phenomena in the sample. In accordance with the theoretical findings
however, no localization effects are observed.

The theoretical work within this paper paves a clear path for the
future observation of Anderson-like localization in cold atomic samples.

\section{Acknowledgements}

We thank L. Santos, L. Sanchez-Palencia and G.V. Shlyapnikov for
fruitful discussions. We acknowledge support from the Deutsche
Forschungsgemeinschaft (SFB 407, SPP 1116, GK 282, 436 POL), the
ESP Programme QUDEDIS, the  Polish government funds
PBZ-MIN-008/P03/2003 (KS) and 1P03B08328 (2005-08) (JZ).

\end{document}